\documentclass[11pt]{article} 
\def\ds{\displaystyle}

\usepackage{amsmath}
\usepackage{amsfonts}
\usepackage{amssymb}\usepackage{amsmath}
\usepackage{amsfonts}
\usepackage{amssymb}

\begin{document}
\pagestyle{headings}
\renewcommand{\thefootnote}{\alph{footnote}}

\title{Massless Particles in QFT from Algebras without Involution}
\author{Glenn Eric Johnson\\Oak Hill, VA.\footnote{Author is reached at: glenn.e.johnson@gmail.com.}}
\maketitle

{\bf Abstract:} The explicit realizations of quantum field theory (QFT) admitted by a revision to the Wightman axioms for the vacuum expectation values (VEV) of fields includes massless particles when there are four or more spacetime dimensions.

\section{Introduction}


It was demonstrated in [\ref{gej05}] that explicit realizations of quantum field theory (QFT) are admitted by a revision to the Wightman axioms for the vacuum expectation values (VEV) of fields. The expansion of QFT beyond consideration of $*$-algebras achieved realizations of the first principles of quantum mechanics, Poincar\'{e} covariance, positive energy, and microcausality for fields exhibiting interaction. Here the development is extended to include massless particles.

The revised Wightman axioms remain satisfied with the inclusion of massless particles. Massless particles require a different algebra of function sequences and four spacetime dimensions. To achieve continuous linear functionals, divergences excluded for finite masses are reconsidered to enable massless particles. The demonstrations of: a semi-norm for a subalgebra ${\cal B}$ within an enveloping $*$-algebra of function sequences ${\cal A}$; Poincar\'{e} covariance; locality; and spectral support do not depend on whether the constituent particles have finite mass and the original demonstrations [\ref{gej05}] suffice. In the case of massless particles, evaluation of scattering amplitudes succeeds just as in the finite mass case although the LSZ asymptotic states are no longer tempered test functions on mass shells.

This note employs the same notation and constructions as [\ref{gej05}] but with the inclusion of zero masses, $m_\kappa=0$.
 
\subsection{Algebras of test function sequences}

A local, Poincar\'{e} covariant, positive energy supported semi-norm on a subalgebra ${\cal B}$ of a Borchers-Uhlmann algebra ${\cal A}$ defines a QFT. In [\ref{borchers}], the algebra of Schwartz functions was denoted $\Sigma$ and for the constructions, a distinct algebra is designated ${\cal A}$. The Fourier transforms of the components of elements of ${\cal A}$ are energy-momentum functions $\tilde{f}_n((p)_n)_{\kappa_1 \ldots \kappa_n}$ that are the multiple argument versions of the span of products of a tempered test function $\tilde{f}({\bf p}) \in S({\bf R}^{d-1})$ and a multiplier $\tilde{g}(p)$ of test functions from $S({\bf R}^d)$ [\ref{gel2}].\[\tilde{f}_1(p):=\tilde{g}(p) \tilde{f}({\bf p})\in {\cal A}.\]To include massless particles, $\tilde{g}(p)$ and all its derivatives are required to vanish at $E=0$. The points with $E_j=0$ are of no consequence in the case of finite masses since the VEV have no support for vanishing energy. With the inclusion of massless particles, this constraint excludes the points where positive and negative mass shells coincide and derivatives of $\omega_j$ are singular at ${\bf p}_j^2$. $\omega_j^2 ={\bf p}_j^2$. The component functions of ${\cal A}$ are test functions of $({\bf p})_n$ when evaluated on mass shells, $\tilde{f}_n((\pm\omega,{\bf p})_n)_{\kappa_1 \ldots \kappa_n} \in S({\bf R}^{n(d-1)})$ and the neighborhood of the origin is excluded. The constructed generalized functions [\ref{gej05}] have the form\[\int dE\, d{\bf p}\; \delta(E\pm \omega) T({\bf p}) \tilde{f}_1(p)=\int d{\bf p}\; T({\bf p}) \tilde{f}_1(\pm \omega, {\bf p})\]for each of the multiple arguments and components, and $T({\bf p})\in S'({\bf R}^{d-1})$, a generalized function.

Factors of $\omega_j$ are not multipliers of the functions [\ref{gel2}] when massless elementary particles are included. To accommodate massless particles, the functions in the subalgebra ${\cal B}$ are set to vanish for all negative energies while preserving that the functions are test functions on mass shells. ${\cal B}$ is a subset of ${\cal A}$ consisting of functions that vanish on negative energies. For every $f_n \in {\cal A}$, let\begin{equation}\label{b-defn} \tilde{\varphi}[ f_n]((p)_n)_{\kappa_1 \ldots \kappa_n} := \prod_{k=1}^n h(E_k/\beta_k) \tilde{f}_n((p)_n)_{\kappa_1 \ldots \kappa_n}.\end{equation}with $h(E)=0$ for all $E\leq 0$ and the infinitely differentiable $h(E)$ and all of its derivative vanish at $E=0$. $\beta_k>0$. For example,\begin{equation}\label{esupp} h(E)=\left\{ \renewcommand{\arraystretch}{1.25} \begin{array}{ll} \exp\left(-\frac{\ds 1}{\ds E}\right)\qquad &E>0\\ 0 &E\leq 0.\end{array} \right.\end{equation}$\varphi[1] := 1$. ${\cal B}$ is $\varphi[{\cal A}]$ and $\underline{W}(\varphi[\underline{f}])$ is bounded when $\underline{W}(\underline{f})$ is bounded.

LSZ states, used to define scattering amplitudes, appear naturally in the development of [\ref{gej05}]. The LSZ states are based upon\[\tilde{\ell}(p_k)=(\omega_k + E_k) e^{i\omega_k t} \tilde{f}({\bf p}_k).\]Evaluation of the scattering amplitudes including massless particles for the LSZ states succeeds for the particular VEV studied. Scattering amplitudes can be evaluated for LSZ states even though the LSZ states are not elements of ${\cal B}$, and the results are as presented in [\ref{gej05}] with the inclusion of the massless particle descriptions.

There are QFT for every realization for $M(p)$, $D$, $S(A)$ that satisfy that satisfy the conditions in [\ref{gej05}]. Example realizations of $M(p)$, $D$, $S(A)$ for massless models include scalar fields with $M(p)=D=S(A)=1$.

\section{Continuity}

When masses are zero, singularities of $B(p)$ and $\Upsilon(p)$ at $p^2=0$ are no longer excluded from the support of the VEV and the $\omega_j$ used in [\ref{gej05}] to define ${\cal B}$ have singular derivatives at ${\bf p}_j^2=0$. The redefinition of ${\cal A}$ and ${\cal B}$ in (\ref{b-defn}) to exclude $E_j^2={\bf p}_j^2=0$ restores $\omega_j$ as a multiplier on the support of the VEV when $m_{\kappa}=0$. For the finite mass cases, the singularities of $B(p)$ and $\Upsilon(p)$ are excluded by $(-p_i+p_j)^2 \geq (m_{\kappa_1}+m_{\kappa_j})^2 >0$ and the $p^2=0$ singularity is beyond the support of the VEV. 
 
From [\ref{gej05}], the VEV are a sum of terms that are products of free field functionals $W_{o;n}((\xi)_n)$ with $V_{n,m}((\xi)_{n+m})$. $V_{n,m}((\xi)_{n+m})$ consists of products of the higher order connected functions with factors that share no arguments. The higher order connected functions are of the form\begin{equation}\label{expand-f}\renewcommand{\arraystretch}{1.25} \begin{array}{l}
T_n((\xi)_{I_n}) = (2\pi)^d c_n\;\delta(p_{i_1}+\ldots p_{i_n}) \; \left( \delta^-_{i_1}\delta^-_{i_2} \overline{M}_{\kappa_{i_1},\kappa_{i_2}}(p_{i_1}\!-\!p_{i_2}) \right.\ldots \times\\ 
 \qquad \qquad \left. \delta^-_{i_{k-1}}\delta^+_{i_k} \beta_{1-i_{k-1}+i_k} B_{\kappa_{i_{k-1}} \kappa_{i_k}}(-p_{i_{k-1}}\!+\!p_{i_k})\ldots \delta^+_{i_{n-1}}\delta^+_{i_n} M_{\kappa_{i_{n-1}} \kappa_{i_n}}(p_{i_{n-1}}-p_{i_n})\right). \end{array} \end{equation}Indices $I_n=\{i_1,\ldots i_n\}$ are distinct and $n\geq 4$. The terms include factors in any combination of $\delta^- \delta^- \bar{M}$, $\delta^+ \delta^+ M$ and $\delta^- \delta^+ B$ that result in at least two factors of $\delta^-$ and two factors of $\delta^+$. $T_n$ vanishes for odd $n$. The $U_n(p)$ are multipliers and need not be considered for continuity. The $M(p)$ are multinomials in the components of $p$ and remain multipliers of tempered test functions. In Section \ref{genf} it is demonstrated that a singularity of the products of the mass shell and energy-momentum conserving delta functions is summable for $d\geq 4$. As a consequence, the products of the mass shell and energy-momentum conserving delta functions define generalized functions for ${\cal A}$ in four or more dimensions. For finite mass, $d\geq 3$ suffices [\ref{gej05}].

Since the factors of the constructed VEV share no arguments, demonstration that each factor is a generalized function is sufficient to define the product. A demonstration that (\ref{expand-f}) defines a continuous linear functional suffices given that the free field is well defined. Factors (\ref{expand-f}) in $\underline{W}$ with no factors of $B(p)$ are continuous linear functionals, the result of Section \ref{genf} since the $M(p)$ are multipliers. The remaining question is the singularities of $B(p)$ and $\Upsilon(p)$.

$B(p)$ and the Lorentz scalar function $\Upsilon(p)$ are represented [\ref{gej05}]\[\renewcommand{\arraystretch}{1.25} \begin{array}{rl} B(p) &= {\ds \int}ds \;\left( a\delta(s) +{\ds \int} d\mu_1(\lambda)\;  \delta^+(s^2-\lambda)\right)\,M(s)\, e^{-p\,s}\\
 &= a\,M(0) +M(-d/dp)\, {\ds \int} d\mu_1(\lambda)\;  \Delta^+(i p;\lambda)\end{array}\]with $\Delta^+(x;\lambda)$ a Pauli-Jordan function of mass $\sqrt{\lambda}$, $\Delta^+(x;\lambda)=(2\pi)^3i D^+(x)$ in the notation of [\ref{bogo}]. $\Upsilon(p)$ has $M(s)=1$. The only finite Lorentz invariant measure $\delta(s)$ results in constant $B(p)$ and $\Upsilon(p)$. Except for these constant terms, $B(-p_j+p_i)$ and $\Upsilon(-p_j+p_i)$ are at least as singular as $1/(p_j-p_i)^2$ [\ref{bogo},\ref{steinmann}]. These constant terms suffice for a demonstration of the existence of a construction that includes massless particles. A distinct $\Upsilon(p)$ can be attributed to each constituent matrix of $M(p)$ in direct sum compositions.

\subsection{The generalized function $\delta(p_1+\ldots p_n)\, (\delta^-)_k\, (\delta^+)_{k+1,n}$ for massless particles}\label{genf}

These generalized functions of $(p)_n \in {\bf R}^{nd}$ are weighted summations over the surface with energy and momentum conserved, and energies on mass shells. Here the definition of\[\delta(p_1+p_2+\ldots p_n)\, (\delta^-)_k\, (\delta^+)_{k+1,n}\]is revisited from [\ref{gej05}] for cases including $m_{\kappa}=0$. The result is that to include massless particles, $d\geq 4$ for the singularities of the measure on the surface $({\bf p})_n \in {\bf R}^{n(d-1)}$ induced by\begin{equation} \delta(P_k):= \delta(\omega_1 \ldots +\omega_k -\omega_{k+1} \ldots -\omega_n)\;\delta({\bf p}_1\!+\!{\bf p}_2\ldots \!+\!{\bf p}_n)\label{delta}\end{equation}to be summable. For massive particles, the generalized function is defined for $d\geq 3$. With $({\bf p})_n$ constrained to the surface with ${\bf P}_k=0$,\begin{equation}\label{surface} {\bf p}_n=-{\bf p}_1\ldots -{\bf p}_{n-1},\end{equation}summation over the surface that satisfies $P_{k(0)}=0$ defines a generalized function except possibly for points on the surface with a vanishing gradient [\ref{gel1}].

Define $s_j$ by\[P_{k(0)}= \sum_{j=1}^n s_j \omega_j\]or $s_i:=1$ for $i\leq k$, and $s_i:=-1$ for $i>k$. The cases $k=0,n$ have no interesting solutions to $P_{k(0)}=0$ and are not considered below. Then $s_1=-s_n=1$. On the indicated surface, the components of the gradient are\begin{equation} \renewcommand{\arraystretch}{1.25} \frac{\partial P_{k(0)}}{\partial p_{j(\ell)}} = s_j \frac{\ds p_{j(\ell)}}{\ds \omega_j}-s_n\frac{\ds p_{n(\ell)}}{\ds \omega_n} \label{grad}\end{equation}for $j = 1,\ldots n\!-\!1$ and $\ell = 1,\ldots d\!-\!1$. The components of ${\bf p}_j$ are designated $p_{j(1)},p_{j(2)},\ldots p_{j(d-1)}$.

For cases including both nonzero and zero masses, the generalized function (\ref{delta}) is always defined since the gradient (\ref{grad}) never vanishes. When $m_{\kappa_j}=0$,\[{\bf u}_j:=\frac{{\bf p}_j}{\omega_j}=\frac{{\bf p}_j}{\|{\bf p}_j\|}\]is a unit vector $\|{\bf u}_j\|=1$ and when $m_{\kappa_i}>0$,\[{\bf v}_i:=\frac{{\bf p}_i}{\omega_i}=\frac{{\bf p}_i}{\sqrt{m_{\kappa_i}^2+{\bf p}_i^2}}\]are the components of a vector of length strictly less than unity, $\|{\bf v}_i\|<1$. These vectors can never be equal for finite ${\bf p}_j,{\bf p}_i$ and consequently (\ref{grad}) is never satisfied. Only cases with all $m_{\kappa}=0$ or all $m_{\kappa}>0$ can be singular. The case with all $m_{\kappa}>0$ has a summable singularity for $d\geq 3$ [\ref{gej05}].

When all $m_{\kappa}=0$, the rest mass does not set an energy scale. For any solution $({\bf p})_n$ to (\ref{grad}), $\beta ({\bf p})_n$ is also a solution for any real $\beta \neq 0$. With ${\bf p}_j=\omega_j {\bf u}_j$ and ${\bf u}_j$ a unit vector, the condition for a vanishing gradient (\ref{grad}) becomes ${\bf u}_j=s_j{\bf u}_1$ and satisfaction of a vanishing gradient is decoupled from energy conservation, $\sum s_j \omega_j=0$.

A neighborhood $V$ of those points in the surface (\ref{surface}) with a vanishing gradient and satisfying energy conservation is\begin{equation}\label{neigh} {\bf u}_j =s_j {\bf u}_1 +{\bf e}_j\end{equation}for $j= 2,\ldots n\!-\!1$ with $\| {\bf e}_j \| < \epsilon$ arbitrarily small and the ${\bf e}_j$ are constrained to\begin{equation}\label{constraint}{\bf e}_j^2=-2 s_j {\bf u}_1 \cdot {\bf e}_j.\end{equation}The constraints preserve the unit vector lengths and reduce the degrees of freedom in ${\bf e}_j$ by one dimension. With $\theta_j$ the angle defined by the two vectors ${\bf e}_j$ and ${\bf u}_1$, the constraint is that $-2s_j \cos \theta_j=\sqrt{{\bf e}_j^2}$. Points with ${\bf p}_j=0$ implying that $\omega_j=0$ are excluded by the selection of functions in ${\cal A}$.

In the neighborhood $V$ of points with a vanishing gradient and conserved energy, (\ref{neigh}) provides that\[\renewcommand{\arraystretch}{1.25} \begin{array}{rl} {\bf p}_n &=-{\ds \sum_{j=1}^{n-1}}{\bf p}_j\\
 &=-{\ds \sum_{j=1}^{n-1}} s_j\omega_j{\bf u}_1 -{\ds \sum_{j=2}^{n-1}} \omega_j {\bf e}_j\\
 &:= -\omega_n c_n\;{\bf u}_1+\omega_n {\bf e}_n\end{array}\]from $s_n=-1$, and with\begin{equation}\label{error-eq}\omega_n{\bf e}_n := -\sum_{j=2}^{n-1} \omega_j {\bf e}_j \quad\mbox{ and}\qquad \qquad \omega_n c_n:= {\ds \sum_{j=1}^{n-1}} s_j\omega_j.\end{equation}To second order in small quantities\[\renewcommand{\arraystretch}{1.25} \begin{array}{rl} \omega_n &=\sqrt{{\bf p}_n^2}\\
 &=\sqrt{\omega_n^2 (-c_n{\bf u}_1+{\bf e}_n)^2}\\
 &=\omega_n \sqrt{c_n^2-2c_n{\bf u}_1\cdot {\bf e}_n+{\bf e}_n^2}.\end{array}\]Squaring and the constraint (\ref{constraint}) results in\[1 = c_n^2+{\ds \frac{c_n}{\omega_n}} {\ds \sum_{j=2}^{n-1}} s_j \omega_j {\bf e}_j^2+{\bf e}_n^2\]and $c_n=1$ when $({\bf e})_{2,n-1}=0$. From (\ref{error-eq}), $s_n=-1$ and to second order in the $({\bf e})_{2,n-1}$,\[\renewcommand{\arraystretch}{1.25} \begin{array}{rl} P_{k(0)} &=(c_n-1)\,\omega_n\\
 &\approx -\frac{1}{2} {\ds \sum_{j=2}^n} s_j\omega_j {\bf e}_j^2\\
 &:= R^2\; \alpha \end{array}\]with\[R^2:=\sum_{j=2}^{n-1} {\bf e}_j^2.\]This change to polar coordinates has $R$ as the Euclidean length of $({\bf e})_{2,n-1}$. $\alpha$ is independent of $R$ consistent with the approximations in $V$.
 
Consistent with the approximations valid within $V$, the energy conserving delta function factors.\[ \renewcommand{\arraystretch}{1.25} \begin{array}{rl} \delta(P_{k(0)}) &= \delta(R^2 \alpha)\\
 &= \frac{\ds 1}{\ds R^2}\;\delta(\alpha) + \frac{\ds 1}{\ds \alpha}\; \delta(R^2 ).\end{array}\]Since the generalized function is defined for $R>0$ and $\alpha$ is independent of $R$, $\delta(\alpha)$ is defined for $R=0$. Then, (\ref{delta}) defines a generalized function when $R^{-2}$ is locally summable. $d\geq 4$ suffices since the Jacobian for the polar coordinates for $({\bf e})_{2,n-1}$ contributes $R^{(d-2)(n-2)-1}$ and $n\geq 4$. The ${\bf e}_j$ are $d-2$ dimensional as a result of the length preserving constraint (\ref{constraint}). The second term vanishes for $d\geq 4$. Accommodation of massless elementary particles results in the physically satisfying result that $d \geq 4$.

Finally, sufficient conditions to include massless particles in the definition of a Wightman-functional from [\ref{gej05}] as a continuous linear functional are: the revision to ${\cal A}$ and the subalgebra ${\cal B}$, constant $B(p)$ and $\Upsilon(p)$, and $d\geq 4$.

\section*{References}
\begin{enumerate}
\item \label{gej05} G.E.~Johnson, ``Algebras without Involution and Quantum Field Theories'', arXiv:1203. 2705v1 [math-ph], 13 March, 2012.
\item \label{borchers} H.J.~Borchers, ``On the structure of the algebra of field operators'', {\em Nuovo Cimento}, Vol.~24, 1962, p.~214.
\item \label{gel2} I.M.~Gel'fand, and G.E.~Shilov, {\em Generalized Functions}, Vol.~2, trans.~M.D.~Friedman, A.~Feinstein, and C.P.~Peltzer, New York, NY: Academic Press, 1968.
\item \label{bogo} N.N.~Bogolubov, A.A.~Logunov, and I.T.~Todorov, {\em Introduction to Axiomatic Quantum Field Theory}, trans.~by Stephen Fulling and Ludmilla Popova, Reading, MA: W.A.~Benjamin, 1975.
\item \label{steinmann} O. Steinmann, ``Structure of the Two-Point Function'', {\em Journal of Math. Phys.}, Vol. 4, 1963, p. 583.
\item \label{gel1} I.M.~Gelfand, and G.E.~Shilov, {\em Generalized Functions, Vol.~1}, trans.~by E.~Saletan, New York, NY: Academic Press, 1964.
\end{enumerate}
\end{document}